\documentstyle [12pt] {article}
\def\ni{\noindent}

\def\deg{\ifmmode^\circ\else$^\circ$\fi}
\def\solar{\ifmmode_{\mathord\odot}\else$_{\mathord\odot}$\fi}

\def\nott{\ifmmode_\circ\else$_\circ$\fi}

\title{ \Large \bf Magnetic Activity in Thick Accretion
Disks and Associated Observable Phenomena: II. Flux Storage}
\author
{Sydney D'Silva and Sandip K. Chakrabarti\\
Tata Institute of Fundamental Research, Homi Bhabha Road, Colaba,\\
Bombay, 400005, INDIA\\}
\begin{document}
\baselineskip 22pt
\maketitle
\begin{abstract}

In paper I, we have studied the conditions under which
flux tubes are expelled from adiabatic thick accretion disks. In the
present paper, we explore a few other models of thick disks, where
flux tubes could be stored. We show that flux tubes with sufficiently
weak fields are not expelled out if they move adiabatically inside an
isothermal disk; they continue to oscillate around mean equipotential
surfaces inside the disk. If the field in the flux tube is amplified
due to the shear, they are eventually expelled away. We explore a `toy'
model also, where the entropy increase outwards from the center of the
thick disk and find a similar behavior. Flux storage in the disk, as in
the case of the sun, in general, enhances the possibility of sustained
magnetic activity formation of coronae in the chimney region. The
existence of  coronae on the disk surface may explain the short-time
variability in the spectra of Blazars and the emission of energetic
particles from AGNs and Quasars. It may also supply matter
to the cosmic jets through magnetized winds.

\end{abstract}

\noindent Keywords: Black holes --  accretion disks -- magnetohydrodynamics
-- magnetic winds -- corona -- flares -- BL Lac objects -- AGNs -- Quasars

\newpage

\noindent {\large \bf 1. INTRODUCTION}

In an earlier paper (Chakrabarti \& D'Silva 1993; hereafter referred to
as Paper I), we have presented a detailed study of how magnetic flux
tubes, either generated inside a disk or advected along the accreting
flow, are expelled outside the disk. Over a large region of the parameter
space, spanned by the size and the field strength, the flux tubes are seen to
emerge in the chimney region of the thick accretion disk. These flux tubes
emerge in buoyancy timescales, and might be advected along the
jets, which are believed to be originated in the chimney of the disk.
It is long known that most of the
luminosity of the disk is emitted from the chimney. For the first time,
we pointed out that the chimneys could also be magnetically the most active
regions in the disk.

Blazars are long believed to belong to that class of active galaxies where the
jets point at the observers. These objects are known to be
highly variable over the entire electromagnetic spectrum.
Variability time scales vary from $10^2$s to $10^5$s (e.g., Bregman 1988).
A large number of models have been proposed to explain the
cause of variability (see, Paper I for a review). If it is caused
by the magnetic activity at the base of radio jets (Blandford and
K\"onigl 1979), then it is imaginable that the flux tubes
are actually anchored in the chimney region of the thick disks and the
activities are due to shock production by flaring events.

In the case of the sun, flux tubes are believed to be anchored
beneath the convection zone where the temperature gradient
is sub-adiabatic, and sunspots are believed to be formed at the
regions where these flux tubes emerge from the surface (Parker 1979).
This emergence takes place due to
Parker's instability (Parker 1955), which is the common cause
of the formation of sunspots, coronae, prominences, flares and other
magnetic regions on the solar surface.
In this paper, we show that thick accretion disks could also entertain
a very similar scenario where flux tubes can be anchored inside
the disk and the parts of these flux tubes could emerge into the chimney,
making it a region of sustained magnetic activity. So far
no work has been done to {\it show} that flux tubes could be anchored
inside a thick accretion disk. In the present paper,
we explore models of thick disks which render efficient anchoring.
We find that isothermal disks are capable of storing flux tubes
provided the tubes move adiabatically. We also explore
some `toy' models of thick disks, which are in hydrostatic equilibrium,
and at the same time, have entropy rising outward from the center. These
disks can store flux tubes provided the field strength is sufficiently
weak so that even the buoyancy force is counter acted.
In other words, in order to anchor a flux tube of given strength
the local entropy gradient should be sufficiently sub-adiabatic.
In may be noted in this context that Shibata et al. (1989), using a two
dimensional MHD simulation show that the Parker instability of
horizontal magnetic flux sheet could be suppressed inside a two temperature
inversion medium. Such models could be physically realised in a variety
of astrophysical circumstances which we elaborate.
Axisymmetric flux tubes, in the absence of shear
amplification and other dynamo effects, continue to oscillate around mean
equipotential surfaces indefinitely, without ever being expelled
out of the disk. Perturbative effects, such as deviation from axisymmetry
(either of the fields, or temperature or pressure profile)
can easily bring these tubes to the surface and cause magnetic activity
as in the case of the sun (Moreno-Insertis, 1986; Choudhuri, 1990;
D'Silva \& Choudhuri, 1993). In the context of accretion disks,
such perturbations could be caused by
various interchange instabilities (Kaisig, Tajima \& Lovelace, 1992)
or non-axisymmetric shear instabilities (Balbus \& Hawley, 1991, 1992;
Hawley \& Balbus, 1992). Shear amplification of the field
could also enhance the field strength till it is expelled away due
to Parker instability (e.g., Matsumoto et al. 1990; Shibata, Tajima,
Matsumoto 1990).

The `toy' models we explore have a simple entropy
distribution, wherein the entire disk becomes radiative and flux
tubes can be stored, or the entire disk is convective and all flux
tubes get expelled out of the disk. In principle, one could
try out entropy distributions where the disk can
have a radiative core and a convective envelope, as in the sun. This
exercise leads to the most obvious question - `What
governs the entropy distribution in an accretion disk?'. The most
serious problem is our ignorance of viscosity and its role in
controlling the dynamics of the entire disk. For instance, if
accretion was surface phenomenon (Pac\'zinski 1978) in thick
disks, then, naively one would expect that the ratio of gas pressure
to radiation pressure $\beta$ would be small at the surface and
increase inward to the center. Since the entropy function $K$ (described
in Paper I) of thick disks goes roughly as $\beta^{-4/3}$, for
$\beta<<1$, this allows the entropy
to rise from the center to the outer edge of the disk. As we
shall show later, this is the Schwarzchild criterion for convective
stability in thick disks. On the other hand, if accretion is
equatorial (Paczy\'nski and Wiita 1980; Wiita 1982, Abramowicz \& Paczy\'nski
1982, Chakrabarti, Jin \& Arnett 1987), the disk can naively be expected
to be convective. Convective stability also implies stability
towards magnetic buoyancy, provided the field strength is low
enough --- the value depends on the entropy gradient.

The presence of magnetically active coronae on disk surface can not only
explain the variability of Blazars, they can also supply matter to the
jets through magnetic winds. The flares can accelerate particles through
the first order Fermi process as in the case of the sun. Thus a large
fraction of the high energy radiation from AGNs may originate at these
flares. Magnetic flux tubes also preferentially filter charged particles
(e.g. protons, electrons and ions) from those uncharged (e.g., neutron)
and may produce neutron rich core of the tori which, in turn, can produce
neutron rich elements through processes involving rapid capture of neutrons.

The plan of the present paper is the following:
In the next Section, we describe model equations for the thick disks
which we study in this paper. The equations of the dynamics of flux tubes
inside a disk have already been presented in Paper I. Here we just
reproduce the equations and show how the quantities {\it inside}
the flux tubes vary, when the tubes move isothermally or adiabatically.
In \S 3, we persent numerical solutions describing the behavior of flux tubes
in the various thick disks. In \S 4, we briefly describe the astrophysical
implications of our results. Finally, in \S 5, we summarize our paper and make
concluding remarks.

\ni{{\large \bf 2. MATHEMATICAL EQUATIONS}}

We consider axisymmetric flux rings, and study their
dynamics under the thin flux tube approximation, both of which are
described in Paper I. Here we reproduce Eqns. (10)-(12) of Paper I.

\begin{eqnarray}
\ddot{\xi} - \xi\dot{\theta}^2 +
{X\over{(1+X)}}[-\xi\dot{\phi}^2\sin^2\theta -
2\xi\omega\dot{\phi}\sin^2\theta] =\nonumber\\
 = {X\over{(1+X)}}\{{M\over X}[g - \xi\omega^2\sin^2\theta] -
T_{ens}\sin\theta - {D_r\over{\pi\sigma^2\rho_e}}\},
\label{eq:one}
\end{eqnarray}
\begin{eqnarray}
\xi\ddot{\theta} + 2\dot{\xi}\dot{\theta} +
{X\over{(1+X)}}[-\xi\dot{\phi}^2\sin\theta\cos\theta -
2\xi\omega\dot{\phi}\sin\theta\cos\theta] =\nonumber\\
 = - {X\over{(1+X)}}\{{M\over X}\xi\omega^2\sin\theta\cos\theta +
T_{ens}\cos\theta + {D_{\theta}\over{\pi\sigma^2\rho_e}}\},
\label{eq:two}
\end{eqnarray}
and
\begin{eqnarray}
\xi\sin\theta\ddot{\phi} + 2\dot{\xi}\sin\theta(\dot{\phi} +
\omega) + 2\xi\cos\theta\dot{\theta}(\dot{\phi} + \omega) +\nonumber\\
 + l\nott(n-2)(\xi\sin\theta)^{(n-2)}[\dot{\xi}\sin\theta +
\xi\dot{\theta}\cos\theta] = 0.
\label{eq:three}
\end{eqnarray}

\ni where $(\xi,\theta,\phi)$ describes the position of a fluid
element inside the flux tube, and the quantities of drag $D$,
dimensionless angular velocity $\omega$ and $X$ are described in
$\S$ 2.2 of Paper I. Magnetic tension is given by

\begin{equation}
T_{ens}={4\pi M\nott T_o\over\mu_e A(1-M\nott)\xi\nott\sin\theta\nott},
\label{eq:T_ens}
\end{equation}

\ni The expansion factor $A=(\sigma/\sigma\nott)^2$ is the ratio of the
cross-sectional area of the flux tube with its initial cross- sectional
area, and can be derived using the mass and flux conservations as

\begin{equation}
A={\rho_e(\xi\nott,\theta\nott)\over \rho_e(\xi,\theta)}
({\xi\nott\sin\theta\nott\over \xi\sin\theta})
{(1-M\nott)\over(1-M)},
\label{eq:A}
\end{equation}
\ni The subscripts $e$ and $i$ refer to the corresponding quantities
outside (ambient medium) and inside the flux tube respectively.
The subscript $o$ refers to the quantities corresponding to the initial
conditions.

\noindent{\bf 2.1 Isothermal thick disks}

In the case when efficient cooling mechanism for ions
is not possible, ions inside the disk may remain very hot, close to the virial
temperature. Such disks have been invoked in the context of two temperature
tori (Lightman and Eardley, 1974) and ion pressure supported tori
(Rees et al. 1982). For the purpose of the present analysis, we
consider isothermal disks, as another extreme from the adiabatic case
discussed in Paper I. Gas pressure obeys the equation of state,
$p_g=R\rho T_d/\mu$,
$T_d$ being the constant temperature of the disk. The radiation pressure
$p_r=(1/3) a T_d^4$ remains a
constant. The equipotential surfaces follow the
same equation as for adiabatic disks (Eqn. 2 of Paper I). The
density variation of the disk obeys the distribution,
\begin{equation}
\rho(\xi) = \rho(R\nott){\rm exp}\{-{\mu\over
RT_d}[{c^2\over 2}({1\over{c_r-1}} - {1\over{\xi - 1}}) +
{l\nott^2\over{2n-2}}[c_r^{2n-2} - (\xi\sin\theta)^{2n-2}]\},
\label{eq:isodisk}
\end{equation}

The flux tubes may radiate very efficiently and remain in isothermal
equilibrium with its local surroundings. On the other extreme,
when the cooling is not efficient, the flux tubes can move adiabatically
inside the disk without exchanging any energy. Sections 2.1.1 and 2.1.2 give
the relevant equations for these two cases.
Throughout the calculations we choose an initial entropic condition
in which the flux tube is in thermal equilibrium with its surroundings.
The condition of pressure equilibrium provides the
initial buoyancy factor to be $M\nott = B^2/(8\pi p_e)$.

\ni{\it 2.1.1 When the flux tubes move isothermally}

We now consider the case where the flux tubes are
in thermal equilibrium with its surroundings. This is possible when the
exchange of energy is very efficient. The radiation pressures
inside and outside the tube are equal. Using the mass and the flux
conservation ($B/\rho_i\xi\sin\theta$ = constant) and the
condition of pressure equilibrium, we get a quadratic equation in $M$,
whose solution is of the form

\begin{equation}
M={1\over 2}[b \pm \sqrt{b^2-4}],
\label{eq:Miso}
\end{equation}

\ni where

$$ b= 2+[{\xi\nott\sin\theta\nott\over \xi\sin\theta}]^2
[{\rho_e(\xi\nott,\theta\nott)\over \rho_e(\xi,\theta)}]^2
{(1-M\nott)^2\over M\nott}.
$$

\ni This equation is similar to Eqn. (15) of Paper I. The
solution with a positive sign is unphysical.

\ni{\it 2.1.2 When the flux tubes move adiabatically}

In this case, the entropy of the flux tube remains a constant throughout
its motion inside the disk. Using the mass and flux conservations
along with the condition for pressure equilibrium, we get a polynomial
in $\rho_i/\rho_e$,

\begin{equation}
k_1({\rho_i\over\rho_e})^{4/3} + k_2({\rho_i\over\rho_e})^2 - 1 = 0,
\label{eq:Madiab}
\end{equation}

\ni where

\begin{eqnarray}
 k_1=\beta_e[{(1-\beta_i)\over\beta_i^4}{R^4\over\mu^4})
{3\over a}]^{1/3}{\mu_i\rho_e^{1/3}\over RT_e},\nonumber\\\\
k_2=\beta_e{M\nott\over(1-M\nott)^2}
({\xi\sin\theta\over\xi\nott\sin\theta\nott})^2 {\rho_e\over
\rho_{e,\circ}},
\label{eq:Madiabk1}
\end{eqnarray}

$$
\beta_e={\rho_e\over\rho_e+{(a\mu/3R)}T_{e,\circ}^3},
$$

$$
\beta_i={(1-M\nott)\rho_{e,\circ}\over (1-M\nott)\rho_{e,\circ}
+{(a\mu/ 3R)}T_{e,\circ}^3}.
$$

\ni{\bf 2.2 A toy model for the radiative thick accretion disk}

Here we present a model of a thick accretion disk in which
the entropy is a function of the potential $W$ of the disk. In
the adiabatic disk, entropy is independent of $W$ and in the
isothermal disk $S\propto -log(\rho)$. Presently,
we use a simple functional form, wherein, either the entire disk
remains radiative or the entire disk becomes convective,
depending on whether the Schwarzchild criterion for convective
stability (Landau \& Lifshitz 1959) is satisfied or not. In
thick accretion disks the Schwarzchild criterion for convective
stability can be written in a modified form as

\begin{equation}
{dK\over dW} > 0.
\end{equation}

\ni We take a linear form for the entropy function $K$ as

\begin{equation}
K(W)=K_1+K_2W\; ,
\end{equation}

\ni where $K_1$ and $K_2$ are constants,

\begin{eqnarray}
K_1={K(W_{out})W_c-K(W_c)W_{out}\over W_c-W_{out}}\; ,\nonumber\\\\
K_2={K(W_c)-K(W_{out})\over W_c-W_{out}}\; ,
\label{eq:K_1}
\end{eqnarray}

\ni $W_c=W(R_c,\theta=90\deg)$ is the value of $W$ at the
center of the disk and $W_{out}=W(R_{out},\theta=90\deg)$ is its
value at the outer edge of the disk which is chosen to be
$R_{out}=20r_g$. This dependence of $K$ on $W$ could be through
$\beta$ or through $\mu$ or both. We assume that the composition
of the disk is homogeneous. The local value of the entropy
function $K(W)$ is given by

\begin{equation}
K(W)=[{(1-\beta(W))\over\beta(W)^4}{3\over
a}({R_G^4\over\mu^4})]^{1/3}\; ,
\label{eq:entropy}
\end{equation}

\ni where $\beta(W)$ is the ratio of the gas pressure $p_g$ to
the total pressure $p_t$ on a given equipotential surface, $R_G$
is the gas constant, $a$ is the radiation density constant and
$\mu$ is the mean molecular weight of the gas. According to the
Schwarzschild criterion, if $K_2>0$, the
disk is radiative, and if $K_2<0$, it is convective. $K_2=0$
gives the adiabatic disk. The potential $W$ increases outwards
from the center of the disk, so $(W_c-W_{out})<0$. In Eqn.
(\ref{eq:K_1}), $K_2>0$ for $K(W_{out})>K(W_c)$; the disk is
convectively stable (radiative disk). For a radiation pressure
dominated disk ($\beta<<1$) entropy goes roughly as
$\beta^{-4/3}$. By choosing $\beta(W_c)>\beta(W_{out})$ the
disk becomes radiative (${dK/dW}>0$), or by choosing
$\beta(W_c)<\beta(W_{out})$ it can be made convective (${dK/dW}<0$).

As in Paper I, we choose the polytropic equation of state
$P=K\rho^{\gamma}$ inside the disk, with $K=K(W)$. We assume that
the disk is a radiation pressure dominated geometrically thick
disk. It is also optically thick, photon and matter act as a
single fluid of polytropic index $\gamma = 4/3$. We also assume
that the equation of state is barotropic, $P=P(\rho)$ and the
force balance equation (Eqn. 1 of Paper I) is integrable
$W(P)=-\int{(dP/\rho)}$, and

\begin{equation}
W-W_{out}=-{1\over 2(\xi-1)} - {l\nott^2\over
2n-2}(\xi\sin\theta)^{2n-2} \; ,
\end{equation}

\ni and $\rho$ is given by

\begin{equation}
\rho=\{{1\over K_2}[({K_1+K_2W_{out}\over K_1+K_2W})^{1/4}-1]\}^3 .
\label{eq:raddisk}
\end{equation}

\ni{\it 2.2.1. When the flux tubes move isothermally}

As usual we assume that the initial entropic condition is such
that the flux tube is in thermal equilibrium with its
surroundings. Thus,

$${\rho_{i,\circ}\over\rho_{e,\circ}} = 1-M\nott .
$$

\ni If the flux tubes remain in thermal equilibrium throughout
their journey in the disk, then the pressure balance condition
with the mass and flux conservations give the expression of the
buoyancy factor $M$ to be the solution of a quadratic as in
Eqn. (\ref{eq:Miso}), where the constant $b$ is

$$ b= 2+({\xi\nott\sin\theta\nott\over \xi\sin\theta})^2
({\rho_e(\xi\nott,\theta\nott)\over \rho_e(\xi,\theta)})^2
{(1-M\nott)^2\over M\nott}{\beta_{e,\circ}(1-\beta_e)\over
\beta_e(1-\beta_{e,\circ})}.
$$

\ni{\it 2.2.2. When the flux tubes move adiabatically}

In this case, $\beta_i(\xi,\theta)=\beta_i(\xi\nott,\theta\nott)$. Using the
mass and flux conservations we get an equation for
$\rho_i/\rho_e$ in the similar form as Eqn. (\ref{eq:Madiab}),
where the constants $k_1$ and $k_2$ are given by

\begin{eqnarray}
k_1={1-\beta_{e,\circ}(W)M\nott \over(1-M\nott)^{4/3}}
{K_{e,\circ}\over K_e}\; \nonumber\\\\
k_2={\beta_e(W)^2(1-\beta_{e,\circ})\over
\beta_{e,\nott}(1-\beta_e)} {M\nott\over (1-M\nott)^2}
({\xi\sin\theta\over \xi\nott\sin\theta\nott})^2 ({T_e\over
T_{e,\circ}})^2 .
\label{eq:Mrad}
\end{eqnarray}

\ni{\bf 2.3 Inclusion of effects of shear amplification}

The field strength of the flux tubes can be amplified by convective
motions and differential rotation, and the azimuthal field can
be shown to grow exponentially (Galeev, Rosner \& Vaiana 1979),
with a growth time $\tau \sim{(\xi/v)}$, where
$v$ is the convective velocity. Following Bisnovatyi-Kogan and
Blinnikov (1977), we approximate convective velocities to
be $v\approx\alpha^{1/3} c_s$, where $c_s$ is the sound speed
and is roughly equal to the local infall velocity $v\approx
{(\alpha/\sqrt{\xi-1})}$; $\alpha$ being the viscosity parameter
(Shakura \& Sunyaev 1973).

Assuming that the shear amplification of field inside a thick disk takes
place at the same rate as inside a thin disk, we choose,

\begin{equation}
B=B^{'}exp\{t/\tau\},
\label{eq:shear}
\end{equation}

\ni where $B^{'}$ is the field in the previous time step and $t$ is the
time interval. It is not our intention to develop a detailed picture of the
dynamo process inside the disk. Our present strategy is to
get some insight about the physical processes
that might take place when such an enhancement of the field is turned on.

When the field is amplified, flux inside the tube is no longer conserved,
so the Eqns. (\ref{eq:T_ens}) and (\ref{eq:Mrad}) change accordingly as

\begin{equation}
T_{ens}={4\pi [1-\rho_i/\rho_e] T(\xi,\theta)\over\mu_e A (\rho_i/\rho_e)
\xi\sin\theta},
\label{eq:shearT_ens}
\end{equation}

\ni and

\begin{eqnarray}
k_1={[1-\beta_{e,\circ}(W)M\nott]\over(1-M\nott)^{4/3}}
{K_e^{'}\over K_e}\; \nonumber\\\\
k_2={\beta_e(W)^2(1-\beta_{e}^{'})\over
\beta_{e}^{'}(1-\beta_e)} {[1-(\rho_i^{'}/\rho_e^{'})]\over
(\rho_i^{'}/\rho_e^{'})^2}
({\xi\sin\theta\over \xi^{'}\sin\theta^{'}})^2 ({T_e\over
T_e^{'}})^2 ,
\label{eq:shearMrad}
\end{eqnarray}

\ni where, a prime ($'$) refers to the corresponding quantities at the
previous integration step.

\ni{\large \bf 3. NUMERICAL RESULTS}

We first study the isothermal disks in $\S$ 3.1 and then the
radiative and convective disks in $\S$ 3.2.

\ni{\bf 3.1 Isothermal disks}

Since we are interested in studying the generic behavior of the flux tubes
in isothermal disks, we choose some typical parameters.
We consider an isothermal thin disk described by Eqn. (\ref{eq:isodisk}),
with the boundary condition of $\rho(R_{out})=10^{-15}$ gm cm$^{-3}$. The
disk is around a black hole of mass of $10^6$ M\solar, with its outer
edge $R_{out}$ at $20 r_g$. The temperature of the disk is taken to be
$T_d=5\times 10^9$ \deg K. The
angular momentum distribution $l=l\nott(\xi\sin\theta)^n$, where $l\nott$
and $n$ are positive constants as in Paper I. In $\S$ 3.1,
we study the behavior of the flux tubes when they are in isothermal
equilibrium with their surroundings and $\S$ 3.2 gives the behavior
when they move adiabatically inside the disk.

\ni{\it 3.1.1 When the flux tubes move isothermally}

We take a constant angular momentum disk ($n=0$) with $l\nott=2$.
Flux rings of $M\nott=0.3$ are released close to the equatorial
plane ($\theta=89\deg$) at $\xi= 4, 5, 6, 7$ and $10$ with the
initial conditions of zero velocity, and the trajectories of the
flux tubes are computed by integrating Eqns.
(\ref{eq:one})-(\ref{eq:three}) with the assumption that drag and
accretion are negligibly small ($D\sim 0$ and $u_{acc}\sim 0$). Figure
1a shows the results; it is almost identical to the corresponding
adiabatic case in Paper I, except that the flux tube released at
$\xi=6$ also emerges into the chimney. This interesting behavior is
because magnetic tension (which goes inversely to the radius of
the flux ring) dominates over magnetic buoyancy. Section 3.3 of
Paper I discusses the importance of magnetic tension. Simple analytic work
which involves balancing magnetic tension and the buoyancy effects,
shows that if $T_d>4\times 10^{10}$ \deg K, magnetic tension dominates over
magnetic buoyancy and the flux tubes released at the outer edge of
the disk should also fall into the chimney. Numerically, when all the
effects are taken into account, this happens at somewhat lower temperature,
even at $T_d=10^{10}$ \deg K itself, as shown in Fig. 1b. All the conditions
in Fig. 1b, except the value of $T_d$, are the same as in Fig. 1a.

In order that the self-gravity of the disk does not become important, we chose
disks of very small density, i.e., we choose a very high temperature.
The black hole mass $M_{BH}=10^6$ M\solar $\;$ and the disk of size
of $20 r_g$ with the outer boundary condition of $\rho(R\nott) =
10^{-15}$ gm cm$^{-3}$ restrict the lower limit on the temperature of the
disk to roughly $10^9$ \deg K. On the other hand, if $T_d$
is increased beyond $10^{11}$ \deg K the disk becomes ion pressure
dominated. The disk parameters of $M_{BH}$, $\rho(R\nott)$ and
$R\nott$ that we have chosen, conspire in such a way that $T_d$
has to lie between $10^9$ and $10^{10}$ \deg K in
order to avoid magnetic tension from dominating the dynamics and
the disk becoming self-gravitating. A smaller temperature can be
chosen provided we choose a smaller disk size. When $T_d$ is
below $10^{10}$ \deg K, the behavior of the flux tubes is
similar to that of the tubes which move isothermally in
the adiabatic disk ($\S$ 3.1 of Paper I). However, when flux
tubes move adiabatically in the isothermal disk the behaviour is
remarkably different as discussed in the next subsection.

\ni{\it 3.1.2 When the flux tubes move adiabatically}

We repeat the calculations with parameters as in Fig. 1a, except
that the flux tubes move adiabatically inside the
disk. This is done by calculating the buoyancy factor $M$ using
Eqn. (\ref{eq:Madiab}). To start with, the flux tubes are
buoyant because of the particular initial condition chosen such
that they are in thermal equilibrium with their local
surroundings, but not in hydrostatic equilibrium. The flux tubes
move out in the direction of the
pressure gradient force and in doing so they expand in order to be
in pressure equilibrium with its immediate surroundings. The
reduction in density due to the expansion, however, is not sufficient to
keep its density smaller than its surroundings, and the tube
becomes less buoyant as it rises, till it reaches a point where
the tube is no longer buoyant ($\rho_e=\rho_i$). The rising
flux tube overshoots this point and enters into a region where
it becomes heavier than its surroundings and begins to fall back causing
it to oscillate back and forth. Figure 2a shows the results.
Flux tubes are seen to oscillate about the
equipotential surfaces and in addition they move along it. The
movement along the equipotential surfaces is due to the tension
which pulls the flux tube toward the rotation axis. If $T_d$
is sufficiently low, where tension can be totally neglected, the
flux tube will not move along the equipotential surface, but
just oscillate both radially and vertically. Figure 2b shows the results for a
$n=0.1$ disk. In addition to the tension which acts toward the
rotation axis, there is now a Coriolis force which acts away
from it; these two competing forces determine the fate of the flux tube.
For example, $M\nott = 0.3$ flux tubes released at $\xi=10$ and $14$ revert
back as soon as they cross the dot-dashed curve. This marks the surface
where the component of effective gravity perpendicular to the
rotation axis becomes zero. This is also the surface where the
component of buoyancy perpendicular to the rotation axis changes
sign from being away from it outside this surface to being
toward it inside this surface. If the field is sufficiently
large as for the $M=0.7$ case, then tension makes the flux tube overshoot
so much that buoyancy makes it oscillate about the equatorial plane.

\ni{\bf 3.2 Radiative and convective disks}

Hereafter, we confine ourselves to the study of flux tubes which move
adiabatically. We do not deal with the isothermal case because the generic
behaviour of the flux tubes is similar to that in adiabatic or isothermal
disks. We choose a radiative thick accretion disk (convectively stable)
by choosing $\beta(W_c)=0.0002$ and $\beta(W_{out})=0.0001$ (See
$\S$ 2.2). Equation (\ref{eq:raddisk}) gives the distribution of
$\rho$ in such a disk. Flux rings of $M\nott=0.3$ are released
from $\xi=5,\ 7,\ 10$ and $15$ and $\theta=89\deg$ in a $n=0$ disk
with zero initial velocity. The trajectories are computed by
integrating Eqns. (\ref{eq:one})-(\ref{eq:three}) assuming that
drag and accretion are negligible ($D=0$ and $u_{acc}=0$)
and the flux tubes move adiabatically in the disk (this is
ensured by using Eqn. \ref{eq:Mrad}).
Except for replacing the isothermal disk with the radiative
disk, the conditions are same as in Fig. 2a. The results are
shown in Fig. 3a. The flux tubes oscillate about an equipotential
surface as in an isothermal disk. For any given entropy gradient there
is a maximum $M\nott$ which can be confined in the disk at a given point.
Figure 3a shows that the flux tube released at $\xi\nott=15$
emerges because the oscillation amplitude increases with
$\xi\nott$ (Eqn. \ref{eq:analy} given below) and the amplitude is
so large that it emerges before it can turn back. Here we
present a simple derivation to show how the
oscillation amplitude depends on $M\nott$ and $\xi\nott$.

Section 3.1.2 described, in detail, how and why the flux
tubes oscillate around an equipotential surface, where its
density is equal to the external density.
(On the equipotential
surface, densities will be exactly equal if $M \sim 0$.).
The variation of density with height is described by Eqn. (\ref{eq:Mrad}).
If a small displacement of the flux tube equalises the external
and internal densities, $(\rho_i/\rho_e)=1$, we can
substitute the constants $k_1$ and $k_2$ from Eqn.
(\ref{eq:Mrad}) and approximate Eqn. (\ref{eq:Mrad}) in the limit when
$\beta_e<<1$, as

\begin{equation}
K_e\sim {K_{e,\circ}\over 1-M\nott}\; .
\end{equation}

\ni Expanding $K_e$ by a Taylor series about the point where the flux tube
was released and retaining only the first term, we have

\begin{equation}
K_{e,\circ}+{\partial K\over\partial W}{\partial W\over\partial
x}|_0 dx \sim  {K_{e,\circ}\over 1-M\nott}\; ,
\end{equation}

\ni which on simplification gives the amplitude of oscillation as

\begin{equation}
dx={K_eM\nott\over (1-M\nott)}{1\over K_2{(\partial W /\partial
x)}} \; ,
\label{eq:analy}
\end{equation}

\ni For $M\nott<<1$, the amplitude of oscillations increases
with $M\nott$ which measures the initial departure
from hydrostatic equilibrium. Beyond the dot-dashed curve (cf. Fig. 2b),
$(\partial W/\partial x)>0$ indicating that the initial displacement of the
flux tube is outward in the direction of the pressure gradient
force. Similarly, between the inner edge of the disk and the
center $(\partial W/\partial x)<0$, hence the initial
displacement of the flux tube is toward the rotation axis
along the pressure gradient force. The constant $K_2=\partial
K/\partial W$ indicates whether the disk is adiabatic, convective or radiative.
If $K_2=0$, the disk is adiabatic, and the initial displacement is infinite;
in other words, the flux tube emerges out in the direction of the pressure
gradient force (Fig. 8 of Paper I). If $K_2>0$ the disk is radiative and the
displacement, or oscillation amplitude, goes inversely to it. The derivation
is not valid for $K_2 <0$. When $\beta(W_c)$ is made even slightly greater than
$\beta(W_{out})$ flux tubes are expelled out as in the convective disks.

\ni{\bf 3.3 Inclusion of the effects of shear amplification}

We study the behavior of the flux tubes that are anchored in a
radiative disk under shear amplification. This is done by amplifying
the field through Eqn. (\ref{eq:shear}) and
substituting Eqn. (\ref{eq:shearT_ens}) for the tension and Eqn.
(\ref{eq:shearMrad}) for the buoyancy factor, while numerically
integrating Eqns. (\ref{eq:one})-(\ref{eq:three}). Figure 4 shows
the results under similar conditions as Fig. 3a; the amplitude of
the oscillations increases and also the mean surface
about which the oscillations take place keeps shifting outward till
the flux tube emerges. The process takes several dynamical timescales
before the flux tubes emerge.

\noindent {\large \bf 4. IMPLICATIONS OF MAGNETIC ACTIVITY INSIDE
THICK ACCRETION DISKS}

The magnetic field brought by a flux tube
depends upon the point of release, the degree to which
the field is saturated in comparison to the equipartition value,
as well as the location where it surfaces. Because of the uncertainties,
we have refrained from providing any quantitative description of the field
itself. In general, for typical values
of the parameters, the field is found to range between a few thousand
Gauss at the inner edge of the disk to a few Gauss at the outer edge.
This is quite strong and
a large number of important observational effects are expected if the
chimneys of the accretion disks do harbour such fields in the form
of loops, coronae and prominences. In the case of Blazars
(i.e., the optically violent variables and the BL Lac objects), the effects
would be more prominent since the jets are supposed to be beamed
directly towards the observers. Indeed, most of the multifrequency
observations of these objects do invoke the presence of very strong
magnetic flares for a satisfactory explanation of the magnitude and
the time scale of variabilities as well as the energies associated with them.
Recent observation of TeV gamma rays from a highly variable source MK421
may also be explained through repeated accelerations in magnetic loops. In the
case of other radio sources, which are not beamed towards the observer, the
presence of magnetic fields in the chimney region can only be inferred
through indirect means, by comparing the kinematic luminosity of cosmic jets
with the mass loss through magnetic winds from the chimney surface,
for instance. Similarly, if significant magnetic flux tubes are present
inside the disks, they would play an important role in influencing
nuclear compositions both inside the disk, as well as outside.
Below, we discuss some of these suggestive effects.

\noindent (a) {\it X-ray flickering and optical microvariabilities in AGN}:

Emission from Blazars exhibit high polarization and variability at all
wavelengths (e.g., Brown et al. 1989) on time scales varying from
roughly months (quiescent emission component) at wavelengths longer than 1 cm
to weeks at submillimeter wavelengths to roughly days (flaring component)
in UV and smaller wavelengths. X-rays can also vary between days
to months. It is believed that the emission originates from a single compact
region (Gear et al. 1986) of size $r\sim 10^{15}$ to $10^{17}$cm and the
magnetic field, which is largely turbulent (Jones at al. 1985)
is $B\sim 0.1 - 1$ G (Gear et al. 1985, Brown et al. 1989).
A large number of these objects such as, OJ287, 3C 345, BL Lac, and 3C 454.3
etc. show clear evidence of a net magnetic field aligned orthogonal to the
direction of the jet axis.
The spectral index $\alpha$ shows a break at a frequency $\nu_b$ which is
known to evolve with time as the emission mechanism changes.
At low frequencies, where the loss during the acceleration is
not important, $\alpha \sim -0.75$ but at a break frequency,
$\nu_b$, when the electrons experience significant radiative loss during
acceleration, the spectrum steepens to have $\alpha \sim -1.25$.
But if the electrons experience a burst of injection
and then the radiative loss, the spectrum should have a slope of
$\alpha \sim -2$ (Khardashev 1962).
This variation is indeed observed. For an efficient radiative loss which
produces steepening at infrared and which is clearly observed, one requires
the field be very high.  All these points to the evidence that a strong
magnetic activity, very similar to what happens of the solar surface,
must be taking place at the chimney wall as well.
Blandford and K\"onigl (1979) suggest that
if a succession of mild shocks continuously reaccelerate the electrons,
the flow should behave as roughly isothermal, and the field perpendicular
to the flow axis at the shock should vary as $B \propto R^{-1}$. This
condition produces a very high field of around $1$G at the flaring region.
In a highly variable source, such as MK421 ($t_{var} \sim 10000$s),
(Brodie, Bowyer \& Tennant 1987) field varies as $B=B_0 r^{-1.4}$
with $B\sim 0.2$G. In PKS2155-304, field varies as $B=B_0 r^{-1.3}$
with $B_0 \sim 100$G close to the core to $70\ \mu G$ at 1pc. The trends of
these observational results agrees with our calculated field values near
the chimney.

In the Introduction of Paper I, we mentioned some models to
account for intraday variability of the blazars. We propose here that
magnetic flares in the chimney can be considered to be a major contributor
to this variability. Though the non-linear physics of the flares on the
solar surface is far from known, it is very reasonable to assume that the
reconnection of the abundant flux tubes in the chimney should produce
flares in time-scales ($t_f$) of the order of a few
times the infall time-scale $t_i\sim \alpha^{-1} r_c/v_r$, since the
flux tubes typically saturate to equipartition value in about $t_i$.
Here, $\alpha$ is the viscosity parameter,
$r_c$ is the center of the disk and $v_r\sim (2GM/r)^{1/2}$.
For a disk around a black hole of mass $10^8M_\odot$, $r_c\sim
10 r_g$ and $\alpha\sim 0.01$, $t_f\sim 1-3$d. The energy deposited
at each flare is about $L^3B_f^2/4\pi$. Here, $L$ is the
length scale of a typical flare and $B_f$ is the average field
on the chimney surface. The energy is released in time $L/V_A$, where,
$V_A$ is the Alfven velocity $(B_f^2/4\pi\rho_f)^{1/2}$, $\rho_f$
being the density of matter inside the chimney. With reasonable estimates
of $L\sim r_g$, $B\sim 100$G, and $\rho \sim 10^{-14}$gm cm$^{-3}$,
one obtains the energy release, time-scale and the rate of release to be
given by $2 \times 10^{37}$ergs, $0.3$d and $2 \times 10^{34}$ erg s$^{-1}$,
respectively. Depending upon the size of the flares their occurence rate
will change. Thus a large number of frequent micro- and miniflares are expected
inside the chimney. Toward the outer edge of the disk, flares will be
very less energetic if frequent, or highly energetic but very less frequent.
Away from the hole, the flare timescale becomes longer because both the
amplification as well as the buoyancy time scales
should grow as $r^{3/2}$ and naively the flare time scales
should also grow as $r^{3/2}$. (The unknown non-linear physical processes
which govern flarings may change this dependence significantly.)
This property explains the rapid flickering in X-rays and
slower variations in optical and radio emissions.

\noindent (b) {\it High energy gamma rays:}

Recently $0.5$TeV gamma rays have been detected from a highly active and
variable source Mk421 (Punch et al. 1992).
It is possible to produce such energetic gamma rays,
if one assumes that they are originated from first order Fermi accelerated
protons radiating at the `thick targets' close to the base of the magnetic
field loops, similar to what is believed to be the source of hard X-rays
and gamma rays on the surface of the sun. However, in the chimney region
where crowded coronae can be present, the particles may be energized
in multiple steps, and not just only in two steps as in the solar corona
(Bai et al. 1983). In this process, moderately energetic protons released in a
flux tube, traverse back and forth and are
repeatedly accelerated at each encounter with shock fronts moving
parallel to the field lines from each footpoint.

In the first order Fermi acceleration, the energy gain is proportional to the
total energy of the particle. Thus protons are more efficiently accelerated
than the electrons for a given velocity. The energy gain per encounter
can be written as (Jokipii 1966):
$$
\Delta W= 2 \beta_s \gamma_s^2 W (\beta_p \mu_p +\beta_s)
$$
where $\beta_s$ is the shock velocity component parallel to the field lines
and $\beta_p$ is the speed of the particle in units of velocity of light,
$W$ is the total energy of the particle, $\mu_p$  is the cosine of the
pitch angle and $\gamma_s$ is the Lorentz factor of the shock motion.
The average time interval between two successive collisions with the
approaching shock fronts is given by (Bai et al. 1983),
$$
\Delta t= \frac {d}{c \beta_p <\mu_p>}
$$
where $d$ is the half length of the flux tube.
Assuming shocks move non-relativistically, i.e., $\beta_s << \beta_p$,
and using the relativistic relation $m=m_p/(1-\beta_p)^{1/2}$, where,
$m_p$ is the rest mass of the proton, one obtains the time variation
of total energy during $i$-th step as,
$$
W_i^2(t)-W_p^2=(W_{i-1}(t_i)^2-W_p^2) exp [4 <\mu_p^2> (v_s/d) (t-t_i)]
$$
where, $W_p=m_p c^2$. Here, $W_{i-1}(t_i)$ is the injection
energy of the proton at the begining of the $i$-th step  at time
$t_i$ (i.e., the final energy of the proton at the end of $(i-1)$-th step).
It is easy to see that in the non-relativistic limit, assuming the duration
of acceleration to be equal to the the shock traversal time of the
tube $d/v_s$ and $<\mu_p^2>=0.75$, the energy goes up by a factor of
$exp(3) \sim 20$ as in the solar flux tubes (Bai et al. 1983).
However, as the total energy goes up, this efficiency factor decreases to
about $exp (1.5) \sim 4.5$ when the ultra-relativistic equation (given above)
is used. Assuming initial injection energy $E_{ini}$
to be about $1$MeV (which is just above the energy where
the gain rate is higher than the Coulomb energy loss rate),
a final energy $W_n$ of $1-2$TeV is obtained in $n$-steps where,
$$
W_n^2 \sim 2W_p E_{ini} exp(3n)
$$
which yields, $n\sim 7$. Because a large number of steps are required,
it may seem to be an inefficient process. But the chimney is more confined
than an almost flat solar surface, and such multiple steps could be
possible. Thus, in general, production of high energy cosmic
rays in the chimney region cannot be ruled out.

One way to verify if this is indeed what takes place in Mk421,
TeV gamma rays can be observed after a delay of several days
or even weeks after the detection of $1-10$MeV gamma rays,
as a considerable time is spent during the acceleration process.
(Indeed, at each step, $d/v_s \sim 2GM/v_s c^2 \sim 3$d
is required, where, $d\sim r_g$ and $v_s\sim 10^8$cm s$^{-1}$ are used.)
Furthermore, intermediate energy photons should also be observed. Note that
the acceleration within shock traversal time does not put any constraint
on the size of the magnetic flux tube. But a delay between different bands, if
observed, would constrain the sizes involved.

\noindent (c) {\it Correlations between radio and optical variabilities}

In the case of BL Lac 0716+714, it has been observed that variabilities
in radio and optical are well correlated (Wagner et al.\ 1990;
Quirrenbach et al.\ 1991) and some kind of quasi periodicities are
observed in time scales of $1.2$d in the first week and of $\sim 7$d
in later weeks of observations. This behavior could be understood
using our observation of the fact that strength of magnetic fields on the
disk surface is well correlated with the pressure inside the disk
where the flux tube is originated. If, for example, the optical
variabilities are caused by the formation and disappearance
of the hot-spots (non-axisymmetric density perturbations) inside the disk
(Chakrabarti \& Wiita 1993), then the radio flares caused by magnetic fields
anchored with these density waves should also follow a similar variability.
A possible delay, corresponding to the buoyancy timescale of a $\leq$day
(inner disk) to several days (outer disk) is expected.

\noindent (d) {\it Matter supply to cosmic jets via magnetic winds}

Magnetic winds carry almost $10^{-14}M_\odot$ yr$^{-1}$
from the surface of the Sun. The active surface of the chimney of
a thick disk, which may have a typical area ($\sim 2 \pi r_c h$, $h$ being the
maxium height of the disk) of $A_f \sim 1000r_g^2$ will release roughly ${\dot
M_w} \sim A_f \rho_f V_E$ amount of matter per second, where $V_E$
is the escape velocity of matter. At an average
distance of $r\sim 10 r_g$, and with $\rho_f=10^{-14}$gm
cm$^{-3}$, the rate will be $\sim M_\odot$ yr$^{-1}$. This is comparable to
the typical kinematic flux observed in a jet. Thus, the magnetic winds from
the chimney can be responsible for supplying necessary matter in the jet.

\noindent (e) {\it Formation of neutron disks and neutron rich isotopes}

In the presence of a large number of flux tubes, which are originated
deep inside, the nuclear composition of the disk could change in buoyancy
time scale. In the case of ion pressure supported thick disks (Rees et al
1982), the central temperature of the disk could be high enough ($\geq
10^{10}$K) that the accreting matter would be photo-dissociated completely
(Chakrabarti, Jin and Arnett 1987; Hogan and Applegate
1987). However, the buoyant flux tubes will carry with them only the
protons and electrons. Neutrons left behind can produce a neutron
rich disk close to the center. Newly accreting matter
can capture these neutrons and produce neutron-rich nuclear isotopes
which eventually escape from the disk by the rising and flaring flux tubes.

\noindent{\large {\bf 5. CONCLUSIONS}}

In the present paper, we have been able to show that it is possible to
construct physical models of thick accretion disks around black holes
which can act as a store house of magnetic fields. In isothermal disks
or in radiative disks, weaker flux tubes
are not expelled away. Instead, they continue to oscillate
around equipotential surfaces, dictated by the forces of gravity and
the centrifugal motion, till they are amplified by shear and
become more buoyant. It is well known that in the case of the sun,
the flux tubes are stored in the regions near the boundary between the
radiative and the convective zones, and appear on the surface due to
instabilities. Similarly, the anchored flux tubes in the disk
are also expected to appear close to the chimney due to
perturbative effects and cause magnetic activities.
We have discussed briefly that a large
number of astrophysical processes, such as supply of matter through
magnetic winds, variability of Blazars through flaring events, production of
highly energetic particles in the coronae through Fermi acceleration
processes, etc. could be related to the interesting
magnetic flux tube behavior that we discover.

In our work, in both Paper I and Paper II, we ignored the effects of
reconnection of flux tubes within the disk. Reconnection takes place
in diffusion timescale, and is slower compared to the shear amplification
timescale. Therefore, it is not expected that it would
have any important effect on the dynamics of flux tubes {\it inside} the
disk. However, outside the disk, in the magnetosphere, the Alfven velocity
is very high (since the matter density is low) and thus reconnection
is important. Indeed, the flares on the solar surface are believed
to be due to this reconnection process.

Though the toy model of the disk we had used is sufficiently realistic
under the circumstances mentioned in the text, it is desirable to
construct a disk model which is fully self-consistent, namely, in which
the structure of the disk is determined from a minimum number of parameters.
For example, we have ignored the effects of nuclear reaction and
associated energy generation inside a hot disk. In Paper I, we have
already shown how various parameters that we use are closely linked with
the `unknown' viscous mechanism operating inside a disk. There is
also a large uncertainty in the size and field distribution of the flux
tubes of the accreted matter. Though, once they are inside the disk they
are amplified to the local equipartition value in a dynamical timescale.
It would therefore appear that a more complete knowledge of the
behavior of the flux tubes and their influence on the observable
phenomena must await an improvement of our understanding of the uncertainties
mentioned above. On the other hand, since we have chosen very reasonable
sets of parameters, we believe that the major conclusions drawn in our
papers (I \& II) will remain intact.

\newpage

\centerline {REFERENCES}

\ni Bai, T. et al. 1983, ApJ, 267, 433\\
\ni Balbus, S.A. \& Hawley, J.F., 1991, ApJ, 376, 214\\
\ni Balbus, S.A. \& Hawley, J.F., 1992, ApJ, 400, 610\\
\ni Balbus, S.A. \& Hawley, J.F., 1992, ApJ, 400, 620\\
\ni Bisnovatyi-Kogan, G.S. and Blinnikov, 1977, A \& A, 59, 111\\
\ni Blandford, R.D. \& K\"onigl, A. 1979, ApJ, 232, 34\\
\ni Bregman, J.N., 1988, in Supermassive Black Holes, ed. M. Kafatos (Cambridge
University Press), p. 43\\
\ni Brodie J., Bowyer S. and Tennant A.: 1987, ApJ, 318, 175\\
\ni Brown et al. 1989, ApJ, 340, 129\\
\ni Brown, L.M., Robson, E.I., Gear, W.K. \& Smith, M.G. 1989, ApJ, 340, 150\\
\ni Chakrabarti, S.K. \& D'Silva, S. 1994, ApJ, (to appear)\\
\ni Chakrabarti, S.K., Jin, L. \& Arnett, W.D. 1987, ApJ, 336, 572\\
\ni Chakrabarti, S.K. \& Wiita, P.J. 1992, ApJ (submitted)\\
\ni Choudhuri, A. R. 1990, ApJ, 355, 733\\
\ni D'Silva, S. \& Choudhuri, A.R., 1993, A\&A, 272, 621\\
\ni Galeev, A. A., Rosner, R. \& Vaiana, G. S., 1979, ApJ, 229, 318\\
\ni Gear, W.K. et al. 1985, ApJ, 291, 511\\
\ni Gear, W.K. et al. 1986, ApJ, 304, 295\\
\ni Hawley, J. F. \& Balbus, S. A. 1991, 376, 223\\
\ni Hogan, C. \& Applegate, J. 1987, Nat, 330, 236\\
\ni Jokipii, J.R., 1966, ApJ, 143, 961\\
\ni Jones, T.W. et al. 1985, ApJ, 290, 627\\
\ni Khardashev, N.S. 1962, Sov. Astr.-- AJ, 6, 317\\
\ni Kaisig, M., Tajima, T. \& Lovelace, R.V.E., 1992, ApJ, 386, 83\\
\ni Lawrence, A., Watson, M.G., Pounds, K.A., \& Elvis, M. 1987, Nat, 325,
694\\
\ni Landau, L.D. and Lifshitz, F.D. 1959, Fluid \ Mechanics, Pergamon
Press (New York)\\
\ni Lightman, A.P. \& Eardley, D.M. 1974, ApJ, 187, L1\\
\ni Matsumoto, R., Horiuchi, T., Hanawa, T. \& Shibata, K., 1990, ApJ,
356, 259\\
\ni Moreno-Insertis, F., 1986, A\&A, 166, 291\\
\ni Paczy\'nski B. 1978, Acta. Astron. 28, 91\\
\ni Paczy\'nski, B. \& Abramowicz, M. A. 1982, ApJ, 253, 897\\
\ni Paczy\'nski B. \& Wiita, P. 1980, Astron. Ap, 88, 23\\
\ni Parker, E. N., 1955, ApJ, 121, 491\\
\ni Parker, E. N., 1979, Cosmical Magnetic Fields, Oxford University Press\\
\ni Punch, M. et al. 1992, Nat, 358, 477.\\
\ni Quirrenbach, A. et al. 1991, ApJ, 372, L71\\
\ni Rees, M.J., Begelman, M.C., Blandford, R.D, \& Phinney, E.S,
1982, Nat, 295, 17\\
\ni Shakura, N. I. \& Sunyaev, R. A., 1973, A\&A, 24, 337\\
\ni Shibata, K, Tajima, T. \& Matsumoto, R., 1990, ApJ, 350, 295\\
\ni Wagner, S., Sanchez-Pons, F., Quirrenbach, A., \& Witzel, A. 1990, A\&A,
235, L1\\
\ni Wiita, P.J. 1982, ApJ, 256, 666\\

\newpage

\centerline {Figure Captions}

\noindent Fig.\ 1 --- Trajectories of flux tubes in the
$R=\xi\sin\theta$ - $z=\xi\cos\theta$ plane, released
at $\xi= 4,\ 5,\ 6,\ 7,\ 10$ and $\theta=89\deg$, with zero initial
velocity and no drag, under isothermal conditions. The dotted line shows the
outer boundary of the disk. The flux tubes have $M\nott=0.1$, the
disk has $n=0$, and the markers are at time
steps of $100 r_g/c$. (a) The disk temperature $T_d=5\times 10^9$ \deg K.
(b) At $T_d=10^{10}$ \deg K where the tension is the dominent force.

\noindent Fig.\ 2 --- Trajectories as in Fig. 1, except that the
flux tubes move adiabatically and they have $M\nott=0.3$
(to highlight the oscillations). (a) Here, $n=0$ i.e., flux tubes
do not feel any Coriolis force, (b) $n=0.1$ and flux tubes feel
Coriolis force.

\noindent Fig.\ 3 --- Trajectories as in Fig. 2, except that the
flux tubes ($M\nott=0.1$) move adiabatically inside a
{\it radiative} disk; (a) $n=0$, (b) $n=0.1$.

\noindent Fig.\ 4 --- Trajectories of flux tubes released at $\xi=5$
and 7, when the field is shear amplified exponentially. Other parameters
are same as in Fig. 3.

\end{document}